# Room-temperature quantum spin Hall phase in laser-patterned few-layer 1T'- MoS$_2$

Naoki Katsuragawa[1], Mizuki Nishizawa[1], Taketomo Nakamura[2], Taiki Inoue[3], Sahar Pakdel[5,6], Shigeo Maruyama[3], Shingo Katsumoto[2], Juan Jose Palacios[4,5] & Junji Haruyama[1*]

[1]Faculty of Science and Engineering, Aoyama Gakuin University, 5-10-1 Fuchinobe, Sagamihara, Kanagawa 252-5258, Japan.
[2]Institute for Solid State Physics, The University of Tokyo, 5-1-5 Kashiwanoha, Kashiwa, Chiba 277-8581, Japan.
[3]Dept Mechanical Engineering, The University of Tokyo, 7-3-1 Hongo, Bunkyo-ku, Tokyo 113-8656, Japan.
[4]Department of Physics, The University of Texas at Austin, Austin, Texas 78712, USA.
[5]Departamento de Física de la Materia Condensada, Instituto Nicolás Cabrera (INC), and Condensed Matter Physics Center (IFIMAC), Universidad Autónoma de Madrid, E-28049 Madrid, Spain.
[6]Department of Physics and Astronomy, Aarhus University, 8000 Aarhus C, Denmark.

[*]Correspondence author. E-mail: J-haru@ee.aoyama.ac.jp

The quantum-spin-Hall (QSH) phase of 2D topological insulators has attracted increased attention since the onset of 2D materials research. While large bulk gaps with vanishing edge gaps in atomically thin layers have been reported, verifications of the QSH phase by resistance measurements are comparatively few. This is partly due to the poor uniformity of the bulk gap induced by the substrate over a large sample area and/or defects induced by oxidation. Here, we report the observation of the QSH phase at room-temperature in the 1T'-phase of few-layer MoS$_2$ patterned onto the 2H semiconducting phase using low-power and short-time laser beam irradiation. Two different resistance measurements reveal hallmark transport conductance values, ~$e^2/2h$ and $e^2/4h$, as predicted by the theory. Magnetic-field dependence, scanning tunneling spectra, and calculations support the emergence of the room-temperature QSH phase. Although further experimental verification is still desirable, our results provide feasible application to room-temperature topological devices.

**Introduction**

The quantum-spin-Hall (QSH) phase of two-dimensional (2D) topological insulators is characterized by a bulk band gap $\Delta$ and helical edge states along the sample edges, where opposite-spin electrons form a Kramers doublet and counter-propagate under the protection of time reversal symmetry. The first experimental demonstration of the QSH phase was realized in HgTe/(Hg, Cd)Te quantum wells, but only at low temperatures[1-4]. Subsequently, InAs/GaSb quantum wells have been a important stage for research of the QSH phase[5,6]. The prototypical QSH insulator was, however, predicted to be graphene[7], but graphene lacks a sizable spin-orbit coupling (SOC) and various efforts have been made to enhance SOC and facilitate the observation of its QSH phase. Indeed, we realized QSH phases in graphene by decorating it with Bi$_2$Te$_3$ nanoparticles at very small coverage (~3%)[8,9]. However, the operating temperature for the QSH phase was still low (< ~20K) and the $\Delta$ values were small (< ~20meV).

Lately, various (atomically) thin layers are attracting significant attention as candidates to present the QSH phase with a large $\Delta$, actually larger than room temperature: the 1T' structural phase of transition metal dichalcogenide (TMDC) monolayers such as WSe$_2$[10] (see Supplementary Note 1) formed on bi-layer graphene/SiC, with Δ ~ 129 meV[11], bismuthene (a bismuth honeycomb monolayer) on SiC, with Δ ~ 0.8 eV[12], and the layered mineral jacutingaite ($Pt_2HgSe_3$), with Δ ~ 110 meV[13]. The 1T' phase of these materials and other TMDC monolayers is particularly amenable to experimental scrutiny (Supplementary Note 1). In fact, large values of Δ have been observed in scanning tunnel spectroscopy (STS) or angle-resolved photoemission spectroscopy (ARPES) at low temperatures, but no reports of the QSH phase at moderate temperatures, *e.g.*, by observing fractional values of the quantum of resistance ($R_Q = h/Ne^2$, where $h$ is the Planck's constant, N is an integer, and $e$ is the electron charge) exist to date. Only one work reported $R_Q/2$ up to 100K in monolayer $WTe_2$ with a very short channel of less than ~ 100 nm[14].

Some reasons can be considered for the lack of experimental evidences of the QSH phase at high temperature, *e.g.,* (1) poor uniformity of Δ over large sample area due to disorder induced by the substrate (doping or potential fluctuations) (Fig. 1a), (2) influence of oxidation (air exposure), (3) influence of conducting substrates, which can short-circuit the edge states, and/or (4) noisy spectra of STM/STS measured at high temperature. Regarding (1) and (2), the observation of a bulk gap by STS and ARPES gives only partial information. Even if the measured gaps are large, thermally activated charge carriers can overcome small values of Δ found at some regions of a large sample area (see lower panels of Fig. 1a). Thus, helical edge states can be disrupted, particularly at high temperatures[15]. Problem (3) can be minimized by using non-conductive substrates (*e.g.*, $h$BN, $SiO_2$).

In order to overcome such shortcomings, we have recently proposed a method to create thin 1T'-$MoS_2$ (~10 layers) systems. They are patterned and embedded into flakes of the 2H semiconducting phase (on $SiO_2$) by laser beam irradiation with a 1 μm diameter beam[16]. In-plane heat accumulation (~300 °C) yielded by the laser irradiation causes the 2H-1T' phase transition[16,17]. This results in a 1T' phase with sizable bulk gap values of Δ ~35 meV over a large area (~2 μm), which shows traces of the QSH phase up to ~40 K[16]. However, in our previous work we could not optimize the laser beam irradiation conditions to clearly obtain one quantized resistance plateau, least of all at room temperature. We attributed this to the formation of too many 1T'-layers and to an excessive damage done by the high-power laser beam (*e.g.*, ~17 mW) and long irradiation time (*e.g.*, over ~30 seconds).

**Results**
**Sample preparation and characterization**

In this work, we have learned to reduce the power and the irradiation time and to apply it to thinner $MoS_2$ layers (~4 layers), which have been previously mechanically exfoliated from the bulk material and transferred onto a $SiO_2$/Si substrate. This allows now to clearly observe room temperature QSH phase arising from a protected topologically non-trivial mono(or bi)-layer embedded in the few-layer 1T'-$MoS_2$. Layer thicknesses of ~6.5 nm have been confirmed by atomic force microscopy (AFM) and optical microscopy. Optical microscopy and AFM images of a flake with several rectangular patterns created by irradiating with a power as low as 4.6 mW (with fixed laser wavelength ~530 nm) and irradiation time of 20 seconds for each point are shown in Figs. 1b – 1c (see also Supplementary Figure 1). In the optical microscopy images, the color of the irradiated patterns changes to semi-transparent. The cross-sectional AFM profile of the irradiated part reveals a decrease in the thickness of ~3.5 nm (Fig. 1c,d). These observations are consistent with our previous reports and those in multilayer $MoTe_2$, which demonstrated a layer thinning effect caused by the burn-out of surface layers due to the in-plane excess heat accumulation via the laser beam irradiation[16,17] (see Supplementary Note 3). The lower laser power and its shorter irradiation times employed in the present experiments minimizes the thinning effect and the damages to the 1T'-phase crystal, particularly for the inner layers, which are protected by the top surface layer. These are likely to present a uniform high-crystal quality.

Typical Raman spectra are shown in Fig. 1e. For the non-laser-irradiated region (2H phase), the Raman peaks are those expected for multi-layer (> five layers) MoS$_2$, showing the large and characteristic E$_{2g}$ and A$_{1g}$ peaks, while the pattern of the peaks for the laser-irradiated region has definitely changed showing only the peaks for the 1T' phase. This suggests that, despite the low power employed, all layers of the 2H phase have been changed into ~ 4 layers in the 1T' phase. Moreover, the peak height (peak/valley ratio) monotonically increases as the laser power decreases. This indicates a higher quality of the present 1T' crystal layers compared to our previous 1T' crystals formed by the high-power laser (17 mW)[16]. Since the spot size of the laser beam used for the Raman spectroscopy is ~1 μm, the high quality 1T' crystal extends, at least, within the observed ~1 μm-diameter area (Supplementary Note 6).

Photoluminescence signals of the laser-irradiated and non-irradiated parts are shown in Fig. 1f. The PL peaks observed in the non-irradiated parts suggest the presence of energy gaps in the 2H semiconducting phase with a layer number > 5, while no photoluminescence signals are obtained in this wave range from the irradiated part. This, again, indicates that all the 2H-semiconducting layers have been converted into either a metallic phase or the 1T' topological phase with much smaller bulk gaps (*e.g.*, ~40 meV in our previous observation[16]) and that could only be detected in a much larger wave-length range. This result is also consistent with the Raman spectra.

**Resistance measurements for room-temperature QSH phase**

Figure 1g shows the same flake of Fig. 1b contacted with Au/Ti electrodes. The Au/Ti electrodes are in contact across the whole width of the bottom rectangular 1T' region so as to measure charge transport along the edges and/or across the bulk of the 1T' region. Less than a 500 nm-length of the top and bottom parts of the electrodes are on top of 2H large-gap regions. Room-temperature resistance ($R$) measurements as a function of back-gate voltage, $V_{bg}$, for the sample shown in Fig. 1g are shown in Fig. 2 (see Supplementary Note 4,5)[18]. The insets schematically indicate the source-drain and voltage electrodes. The low contact resistance at the metal electrodes/the 1T' region interface is subtracted (see Supplementary Note 4). Peaks or plateau-like features with values ~$R_Q/2$ appear in all the two-probe measurements (Figs. 2a-2c), which is consistent with a previous report in a short-channel monolayer of WTe$_2$[14]. This result is a remarkable improvement upon our previous high-power laser irradiation results where, despite being promising, a single clear plateau was hardly obtained[16].

The room-temperature $R$ peak values observed in Fig. 2a for the 0.5 μm-channel are slightly smaller than the theoretical $R_Q/2$ value. However, they saturate to $R_Q/2$ at T = 1.5K, which strongly suggests the presence of current-carrying helical edge states and the QSH phase along this short channel. In this sense, the slight deviation may be associated with thermally activated bulk charge carriers at T = 300K in the 1T' topological region (Supplementary Note 4). In contrast, the $R$ peak values observed in Fig. 2c, which has been measured for the longest channel length (2 μm), is slightly larger than the exact $R_Q/2$ value (*i.e.*, ~0.1$R_Q$). This deviation qualitatively agrees with the $R$ plateau value observed by Young *et al.*[19], attributed to defective edges (*e.g.*, local charge puddles)[5,9,20]. Indeed, a somewhat irregular boundary exist at some points of our 1T'/2H junction, which may lead to local charge puddles[20]. In contrast to our results, however, We *et al.* reported large deviations from ~ $R_Q/2$ in long-channel samples (>> a few 100nm)[14]. This discrepancy suggests highly uniform bulk gaps over long distances in our devices, arising from the high crystal quality achieved and from the protection of the embedding 2H phase.

This result is further confirmed by a four-probe measurement, as shown in Fig. 2d, where a well-developed plateau with a value of $R_Q/4$ appears. The $R_Q/4$ value becomes even more well-defined at T = 1.5K. This value has been reported in *H*-letter like patterns also using four electrode probes on HgTe/(Hg, Cd)Te quantum wells[3], our Bi$_2$Te$_3$-nanoparticle decorated graphene[8,9], and in our previous high-power laser-irradiated thin 1T' MoS$_2$[16]. A straightforward calculation based on the Landauer-Büttiker formalism confirms the measured value for this contact geometry (see Supplementary Note 10).

As mentioned in the introduction, a plausible scenario is that the top surface layer(s) is damaged by oxidization in air and the bottom layer is disordered by the SiO$_2$ substrate, so they cannot exhibit the topological phase. As reported in our previous publication[16], the stacking of 1T' layers does not change the essential band structure features, namely, the band inversion and the presence of a bulk gap. Therefore, if only one monolayer presents a sufficiently good crystallinity, it should present helical states at the 1T'-2H interface (see actual calculation and discussion below) and be responsible for the quantized transport signature (see Supplementary Note 6). We cannot discard, though, that one more monolayer further away from the surface can also present good enough crystal quality. However, it is unlikely that the electrodes can make contact to it or its helical states be accessible since the 1T' areas of the monolayers are not expected to exactly match (Supplementary Note 6).

Perpendicular magnetic-field ($B_\perp$) dependence measurements for the contact geometries in Fig. 2 are shown in Fig. 3a. As $B_\perp$ increases, the conductance $G$, corresponding to the inverse of $R_Q/2$ (red symbols) and $R_Q/4$ (blue symbols), as observed in Figs. 2a,b, respectively, exponentially decrease. $G$ values (black symbols) corresponding to the inverse of $R$ away from the peak or plateau remain almost unchanged. These results are in good agreement with those observed in WTe$_2$[14] and in our Bi$_2$Te$_3$ decorated graphene[8,9], and further support that the $R$ peaks and plateaus can be attributed to helical edge states. Only when the Fermi level is set to the Kramers degeneracy point at $B_\perp = 0$, $R$ can reflect the time-reversal symmetry breaking and gap opening caused by the applied $B_\perp$ (see two insets for red and blue symbols), resulting in the observed exponential decrease in $G$.

**Low-temperature scanning tunnel spectroscopy**

Examples of low-temperature STS spectra taken on laser irradiated regions are shown in Fig. 3b (see also Supplementary Note 7). For $V_{bg}$ ~35V, approximately corresponding to the Kramers degeneracy point (see schematic views in the insets), STS gaps of $\Delta$ ~ 60 - 70 meV, which are larger than room temperature, are measured in two bulk points (blue and green curves), while the gap closes near an edge (red solid curve). Figure 3c shows a distribution of bulk $\Delta$ values measured along the sample longitudinal direction averaged over three different transversal positions. We see a high uniformity in the values measured, ranging between ~50 – 80 meV over ~10 μm (red symbols), to be compared with the smaller values obtained using 17mW-power laser irradiation (black symbols)[16], which also visibly present a larger dispersion. The $\Delta$ values are also larger than 45-meV, as reported in 1T'-WTe$_2$[14]. Moreover, when $V_{bg}$ is set away from the Kramers degeneracy point ($V_{bg}$ ~ 10V; red dotted curve) at the edge position (refer insets), the d$I$/d$V$ transforms into that expected for just the metallic behavior of the 1T' phase. This further re-enforces our evidence for 2D topological insulating states with a high uniformity in the bulk gap $\Delta$. This result also suggests an improvement of the 1T' crystal quality of the top surface layer, although the damage is still too large to allow for a continuous tip scanning over a large area (Supplementary Note 6). This improvement should also reflect in the inner layers, which are protected by the top surface layer. This further adds to our argument that a highly uniform $\Delta$ in the inner layers leads to the observation of the room-temperature-QSH phase.

**Discussion**

Finally, we present theoretical results that confirm the existence of helical states at the 1T'-2H interface. We have first carried out DFT calculations of a periodic heterostructure of alternating 1T'- 2H strips for increasing widths of both phases until confinement effects disappear from the density in the bulk of the strips (Supplementary Note 9). Spin-orbit coupling has been added a posteriori following Pakdel *et al.*[21]. From these calculations we have obtained all the necessary self-consistent Hamiltonian and overlap matrices with which we can compute the Green's function projected near a single interface (see inset in Fig 4 for an schematic picture of the interface) and thereof the density of states (DOS) as a function of wave vector parallel to the interface. A standard recursive Green's functions methodology has been used to compute the semi-infinite 1T' and 2H bulk electronic structure on both sides of the interface[22]. Fig. 4 shows the computed DOS projected in the vicinity of the interface. One can clearly see the bulk inverted bands, a finite $\Delta$, and, most importantly, the characteristic interface bands of helical states crossing the Fermi energy (in the gap) an odd number of times, as expected for a 2D topological insulator.

In conclusion, we have demonstrated the possibility of improving the crystal quality of few-layer 1T'-$MoS_2$ (particularly the inner layers) patterned on 2H-semiconducting $MoS_2$ by low-power and short-time laser beam irradiation. This has made it possible the observation of quantized resistance plateaus, with values in agreement with theory. Although an alternative and more direct confirmation of single-layer transport is lacking at this point, we have attributed such plateaus to a stable room-temperature QSH phase in one of the layers, which is maintained at least for 10 days in air (see Supplementary Note 8). While further experiments should confirm our somewhat speculative conclusion, the present laser-beam irradiation method brings several advantages into the field of atomically thin QSH crystals and also opens the possibility for room-temperature applications of these topological phases. In particular, the on-demand laser-beam patterning of the topological phases on thin semiconducting $MoS_2$ regions facilitates the design of topological spintronic-devices with zero-dissipation. Combining this with high-temperature superconductors could open the door to Majorana-fermion based topological devices and circuits for high-temperature operation[23,24].

**Methods**

In previous experiments for thin 2H-MoS$_2$ layers[16], seven points on a thin MoS$_2$ flake were irradiated by laser beam (532-nm wavelength, 1-μm diameter) with various irradiation times (see Supplementary Figure 1A; points 1, 2, 3, 4, 5 correspond to 500 s, 50 s, 30 s, 20 s, 10 s, respectively, under a constant power of 50 mW) and different powers (50 mW for points 1-5, for 25 mW with 50 s duration for point 6). Optical microscopy image confirmed that six points became semi-transparent and that only a point with an irradiation time below 10 s under a power below 25 mW showed in no changes in the color.

Some of the semi-transparent points (e.g., by 50 mW for 50 s) were analyzed by Raman spectroscopy using a laser power of 0.82 mW and wavelength of 532 nm and by PL with the same power. The laser-irradiated samples (by 50 mW for 50 s) were also analyzed by XPS.

Based on these previous experiments, some 1T'-phases are patterned onto the present few-layer 2H-MoS$_2$ flakes by lower-power laser beam (4.6-mW power, 532-nm wavelength) irradiation (Fig. 1; the 1T'-rectangular patterns. For the laser beam patterning, we used a laser beam mapping (*i.e.*, laser beam scanning) of photoluminescence facility by optimizing the irradiation steps (*i.e.*, overlapping area of two irradiation points) (Supplementary Figure 1B). For example, we changed the overlapping length from 0.25-0.75 μm with 0.25 μm step and checked the crystal quality by optical microscope, AFM, FESEM, and Raman, and selected the best condition which yields the best crystal quality. We also tried to fabricate various patterns by using this method and confirmed its effectiveness (Supplementary Figure 1C, 1D).

**Data availability**

The data that support the findings of this study are available within the paper and its supplementary information file.

**Acknowledgements**

The authors thank S. Murakami, T. Yamamoto, S. Tarucha, T. Ando, T. Enoki, A. MacDonald, R. Wu, J. Alicea, M. Dresselhaus, P. J.-Herrero, and P. Kim for their technical contributions, fruitful discussions, and encouragement. The work carried out at Aoyama Gakuin University was partly supported by a grant for private universities and a Grant-in-Aid for Scientific Research (JP15K13277) awarded by MEXT. The work at the University of Tokyo was partly supported by Grant-in-Aid for Scientific Research (JP17K05492, JP18H04218, and JP19H00652). JJP and SP acknowledge Spanish MINECO through Grant FIS2016-80434-P, the Fundación Ramón Areces, the María de Maeztu Program for Units of Excellence in R&D (CEX2018-000805), the Comunidad Autónoma de Madrid through Grant No. S2018/NMT-4321 (nanomagCOST-CM), and the European Union Seventh Framework Programme under Graphene Flagship Grant No. 604391. SP was also supported by the VILLUM FONDEN via the Centre of Excellence for Dirac Materials (Grant No. 11744) and acknowledges the computer resources and assistance provided by the Centro de Computación Científica of the Universidad Autónoma de Madrid and computer resources at MareNostrum and the technical support provided by Barcelona Supercomputing Center (FI-2019-2-0007) within the RES.

**Author Contributions**

N.K., K.U., M.N., T.N., and T.I. performed the experiments. S.P. and J.J.P. performed the numerical calculation. S.M., S.K., and J.H. designed the experiments. J.H. analyzed the data. J.J.P. and J.H. wrote the manuscript with comments from all authors.

**Author Information**

Correspondence and requests for materials should be addressed to J.H. (J-haru@ee.aoyama.ac.jp). Reprints and permissions information is available at www.nature.com/reprints.

**Competing financial interests**

The authors declare no competing financial interests.

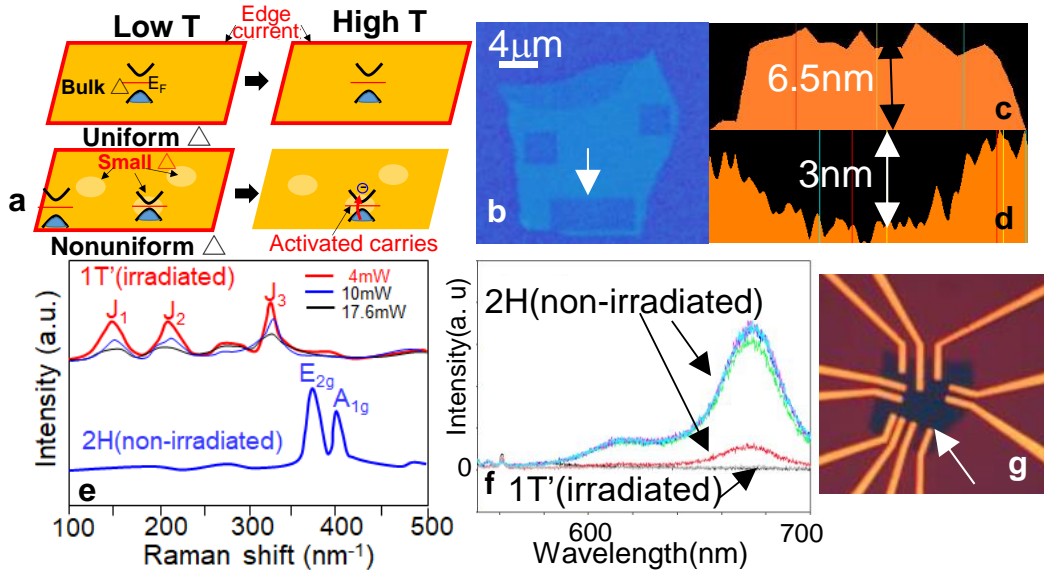

**Fig.1 Device fabrication and characterization. a,** Schematic views of bulk gaps Δ and edge current for uniform and nonuniform cases at low and high temperatures. Thermally activated charge carriers can overcome small Δ located at some parts of the nonuniform sample as temperature increases, obliterating the presence of helical edge states. **b,** Optical and **c,d** cross-sectional atomic force microscope images of a $MoS_2$ flake. The semitransparent rectangle parts in (b) are the low-power laser-beam irradiated regions. The region shown by an arrow is used for the present experiment. **c,** Profile before laser irradiation. **d,** Profile for the laser-irradiated part showing reduction of the layers due to the burn out caused by excess heat accumulation by laser irradiation. It results in the semitransparent parts in (b). **e,** Raman and **f,** Photoluminescence spectra at the partial points for the laser irradiated and non-irradiated regions. The spectra for the irradiated parts in (e) are the results for three different laser powers and $J_1$-$J_3$ peaks unique to the1T' phase grow as laser power decreases, suggesting all the 2H phase layers have been changed into the 1T' phase with a high crystal quality. In (f), no signals are obtained in this wave range from the irradiated part, indicating that all the 2H- phase layers have been converted into either a metallic phase or the 1T' topological phase with much smaller bulk gaps. **g,** Optical microscope image using (b) with Au/Ti electrodes. The region shown by an arrow corresponds to same region in (b), mainly used for the present experiment.

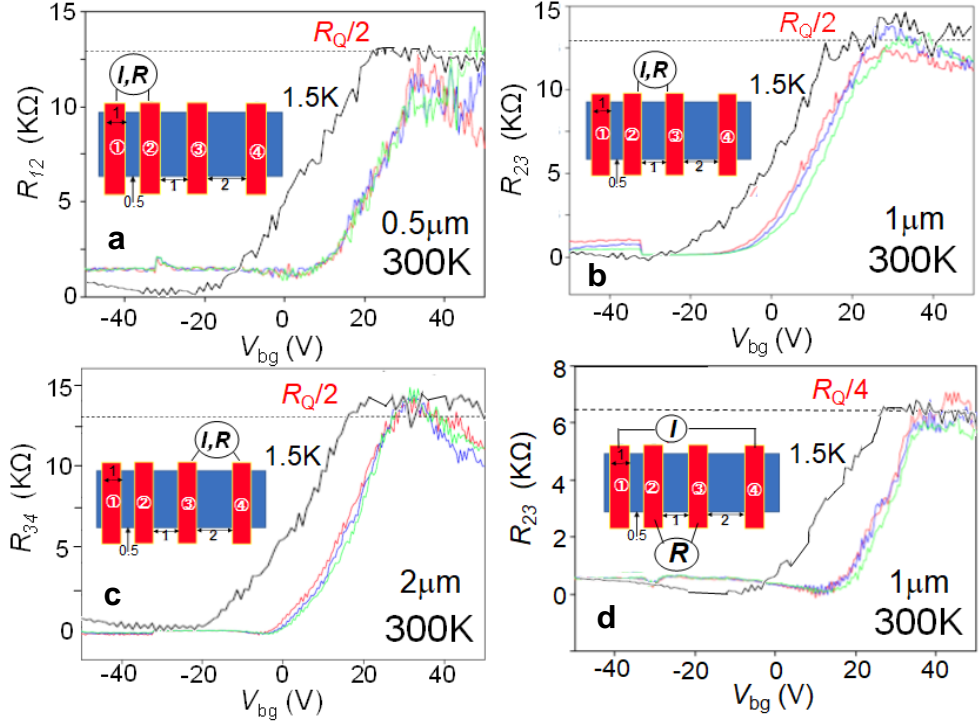

**Fig.2 Room-temperature resistance near $R_Q/2$ and $R_Q/4$. a-c** Two- and **d,** Four-probe resistance (R) measurements as a function of back gate voltage ($V_{bg}$) for the device shown at the bottom of Fig. 1g under a constant current of 1 μA. The low contact resistance at the metal electrodes/the 1T' region interface is subtracted. Different colors of the three curves represent different measurements over time. Dotted lines correspond to $R_Q/2$ and $R_Q/4$, respectively. Black curves are the results measured at 1.5 K. **Insets**: Schematic views of the electrode probe arrangement (red areas) on the 1T' region (in blue, corresponding to the part indicated by an arrow in Fig.1g) with inter-probe (channel) distance expressed in μm. Peaks or plateau-like features with values ~$R_Q/2$ appear in all the two-probe measurements (Figs. 2a-2c), which is consistent with a previous report of QSH phase in a short-channel monolayer of $WTe_2$. This result is further confirmed in Fig. 2d with a value of $R_Q/4$, which has been reported in QSH phase for the *H*-letter like patterns using four electrode probes on HgTe/(Hg, Cd)Te quantum wells, our $Bi_2Te_3$-nanoparticle decorated graphene, and in our previous high-power laser-irradiated thin 1T' $MoS_2$. A straightforward calculation based on the Landauer-Büttiker formalism for Fig. 2d reconfirms presence of the QSH phase.

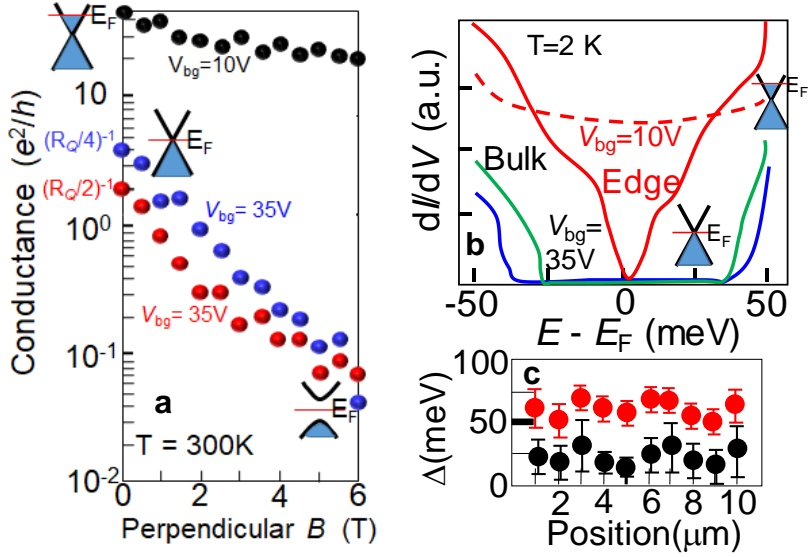

**Fig. 3 Magnetic-field dependence of conductance and STS spectra. a,** Out-of-plane magnetic-field ($B_\perp$) dependence of conductance corresponding to the inverse of the $R$ peak and plateau values in Fig. 2b,d (red and blue symbols) and to the inverse of a $R$ value off peak(plateau) (black symbol). **Insets:** Schematic views of band diagrams near or away from Kramers degeneracy point (KDP) (*i.e.*, crossed points of two lines) with Fermi level ($E_F$; red lines) for each $V_{bg}$. The black symbol remain almost unchanged against the $B_\perp$ increase, suggesting that $E_F$ stays away from KDP (left upper inset). In contrast, the red and blue symbols exponentially decrease with the $B_\perp$ increase, suggesting that $E_F$ at $B_\perp = 0$ stays just at KDP (left lower inset) but a Zeeman gap opens at KDP with increasing $B_\perp$ (right inset). These confirm that the $R$ peak and plateau values in Fig. 2b,d originate from the QSH phase, which appears only at KDPs. **b,** Examples of STS spectra for laser-irradiated 1T' region. The two bulk signals (blue and green curves) were measured near the center of the 1T' region, and the edge signals (red solid and dotted curves) were measured near the boundary of the 1T'/2H phases. $V_{bg}$ was set to +35 V (corresponding to the $R$ peak) and +10V (to off-$R$-peak) for the three solid and the dotted curves, respectively. **Insets:** Schematic views of band diagrams similar to insets of (a). The red dotted curve has non-zero $dI/dV$ values, suggesting the $E_F$ staying away from KDP (upper inset). In contrast, bulk gaps and those close are observed for other three solid curves, suggesting the $E_F$ staying at KDP. These also confirm that the $R$ peak value in Fig. 2b is attributed to the QSH phase. **c,** Distribution of bulk gaps Δ observed by STS on 10 different bulk points along the sample longitudinal direction in the 1T' regions, at $T = 2K$ and $V_{bg}$ ~35V. Red and black symbols correspond to Δ for 1T' regions fabricated by 4.6 mW- and 17 mW-power laser beam irradiation, respectively. Error bars show the results for three different positions (within ~50 nm distance) at the same longitudinal position. The larger Δ values with the smaller distribution in red symbols suggest an improvement of the 1T' crystal quality of the top surface layer and also inner layers by reducing laser power.

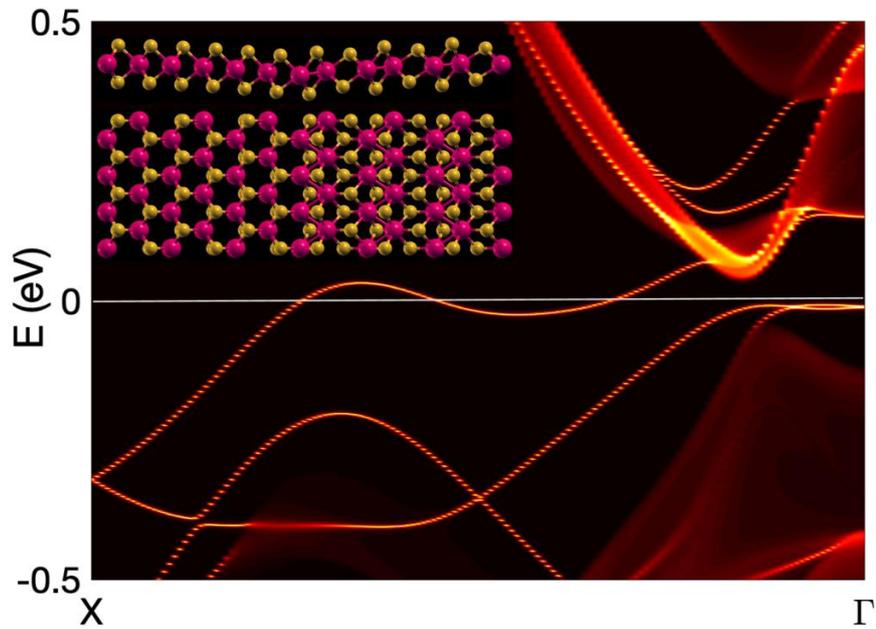

**Fig.4 DFT calculation of a single 1T'-2H interface.** Density of states as function of wave vector parallel to the 1T'-2H interface projected in the vicinity of the interface (shown in the inset). The inverted bulk bands and the bulk gap opened by spin-orbit coupling can be seen near Gamma. The brightness reflects the weight of the states near the interface. The thin lines continuously connecting the conduction band with the valence band correspond to the protected helical states located at the interface. The protection is evident from the fact that the interface band crosses three times the Fermi energy (set to zero) or, in general, an odd number of times if small deviations are allowed within the gap.